\documentclass[journal=jacsat,manuscript=article]{achemso}

\usepackage[version=3]{mhchem} 
\usepackage{lipsum} 
\usepackage{color}
\usepackage{graphicx}
\usepackage{dcolumn}
\usepackage{bm}
\usepackage{bbold}
\usepackage{upgreek}
\usepackage{amsfonts,color}
\usepackage{amsmath}
\author{Hanchen Wang}
\email{hanchen.wang@mat.ethz.ch}
\affiliation{%
Department of Materials, ETH Zurich, Zurich 8093, Switzerland
}%
\author{William Legrand}
\affiliation{%
Department of Materials, ETH Zurich, Zurich 8093, Switzerland
}%
\author{Richard Schlitz}
\affiliation{%
Department of Materials, ETH Zurich, Zurich 8093, Switzerland
}%
\author{Pietro Gambardella}
\email{pietro.gambardella@mat.ethz.ch}
\affiliation{%
Department of Materials, ETH Zurich, Zurich 8093, Switzerland
}%

\title[An \textsf{achemso} demo]
{Current-controlled magnon-magnon coupling in an on-chip cavity resonator}

\abbreviations{IR,NMR,UV}

\begin{document}

\begin{tocentry}

\hspace{-0.5cm}\includegraphics[width=120mm]{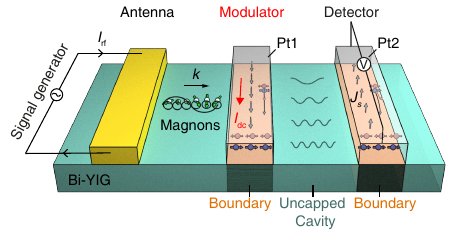}

\end{tocentry}

\begin{abstract}
Harnessing spin currents to control magnon dynamics enables new functionalities in magnonic devices. Here, we demonstrate current-controlled magnon-magnon coupling between cavity and boundary modes in an ultrathin film of Bi-doped yttrium iron garnet (BiYIG). Cavity modes emerge in a BiYIG region between two Pt nanostripes, where interfacial anisotropy modifies the magnon dispersion. These modes hybridize with boundary magnons confined within the Pt-capped BiYIG, resulting in an anti-crossing gap. Modeling based on dipole-exchange spin-wave dispersion accurately reproduces the observed modes and their hybridization. Spin current injection via the spin Hall effect in a Pt nanostripe disrupts the cavity boundary conditions and suppresses both cavity modes and hybridization upon driving the system beyond the damping compensation threshold. Furthermore, tuning the microwave power applied to a microstrip antenna enables controlled detuning of the anticrossing gap. Our findings provide a platform for exploring spin current-magnon interactions and designing on-chip reconfigurable magnonic devices.

\end{abstract}


\newpage

Spin waves, or magnons, are collective excitations of magnetic systems that enable the transport and manipulation of spin angular momentum without involving net charge flow. Due to their wave-like nature and low energy dissipation, magnons hold great promise for applications in information transmission, storage, and processing~\cite{Chumak2015,Pirro2021,Kruglyak2010,Serga2010,Vlaminck2008,han2023coherent,yu2021magnetic,yu2024nonhermitian,brataas2020spin,grassi2022higgs}. 
Realizing functional magnonic devices, however, requires fine control over spin-wave dispersion, coherence, and interaction. Drawing inspiration from photonics and phononics, where nanostructured cavities are routinely used to confine light and sound waves and control their spectral properties~\cite{vahala2003optical,aspelmeyer2014cavity}, recent efforts have aimed at creating analogous confinement structures in magnonic systems. Such magnonic cavities not only enable quantization of spin-wave modes but also provide a versatile platform to engineer magnon-magnon interactions and enable coupling to photons and phonons in hybrid architectures~\cite{Tabuchi2015,guo2023strong,An2020,Schlitz2022,bialek2023cavity,pishehvar2025ondemand}. A variety of approaches have been proposed to implement on-chip magnon confinement, including interfacial exchange coupling~\cite{klingler2018spin,qin2018exchange,liensberger2019exchange}, interlayer dipolar coupling~\cite{chen2018strong, santos2023magnon, qin2021nanoscale, vogel2015optically, hong2021tunable}, magnon traps~\cite{yu2020magnon, chen2021reconfigurable}, and proximity to superconducting materials~\cite{yu2022efficient, borst2023observation}. While these approaches mark significant progress, most existing magnonic cavities are static in nature, lacking the dynamic tunability and reconfigurability that are essential for developing programmable and responsive spintronic systems. A major challenge, therefore, is to introduce mechanisms that allow for in-situ control of cavity formation and mode coupling, ideally through electrical stimuli.

Electrically generated spin currents via charge-spin conversion in nonmagnetic conductors like Pt enable magnetization control and modification of the magnon population in adjacent magnetic materials through spin angular momentum transfer~\cite{manchon2019current, demidov2017magnetization}. Besides magnetization switching~\cite{miron2011perpendicular, avci2017current}, spin currents enable efficient detection and injection of magnons in thin films of low-damping magnetic insulators such as yttrium iron garnet (YIG)~\cite{kajiwara2010transmission,goennenwein2015nonlocal,cornelissen2015long,padron2011amplification,pirro2014spinwave,sklenar2015driving,evelt2016high,collet2016generation,demidov2020spin,uchida2010observation,saitoh2006conversion,noel2025nonlinear}. Recently, progress has been made to manipulate magnon modes in spintronic systems via spin-orbit torques (SOTs) induced by spin current injection and absorption. Notable examples include SOT-driven damping compensation to enhance incoherent magnon conductivity~\cite{wimmer2019spin,cornelissen2018spin}, amplification of propagating spin waves~\cite{merbouche2024true}, and dynamic tuning of parametric pumping~\cite{yang2022parametric}. Integrating SOTs into magnon cavities would additionally offer on-chip tunability and reconfigurability, as required to provide advanced functionalities in magnonic devices~\cite{zhao2023control, rai2023control}.

\begin{figure}
\hspace{-0.5cm}\includegraphics[width=120mm]{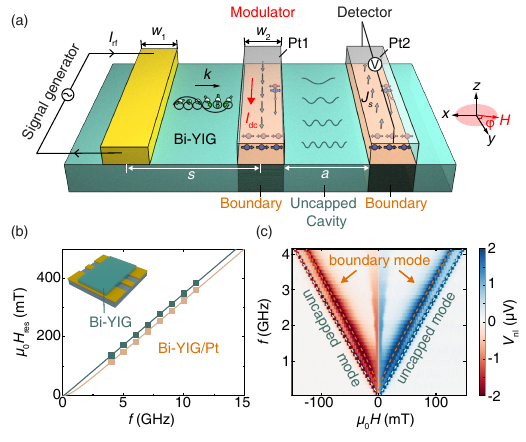}
\caption{(a) Scheme of the cavity and nonlocal spin pumping device. (b) FMR field versus microwave frequency obtained from 3~nm-thick BiYIG blanket layers, with and without Pt capping. The solid lines are the fits based on Kittel's formula. The inset shows the FMR measurement configuration. (c) Nonlocal spin pumping voltage spectra measured in the Pt2 nanostripe, in which the modes of the uncapped and capped (boundary) BiYIG regions are observed.}
\label{fig1}
\end{figure}

In this work, we demonstrate the control of magnon-magnon coupling between on-chip cavity and boundary modes in a 3-nm-thick magnetic insulator, using spin-polarized currents. The magnonic cavity is created by confinement between two boundary areas capped with thin Pt nanostripes [Fig.~\ref{fig1}(a)], which play several important roles: they induce easy-plane interfacial magnetic anisotropy in the magnetic system, shifting up the magnon frequencies and causing magnon reflection at the boundaries~\cite{lee2023large,lee2020interfacial}; they function as inverse spin Hall detectors in sensing the cavity modes and their coupling with the boundary mode through nonlocal spin pumping~\cite{wang2024broad}; they enable control of the magnon modes by spin-polarized current injection. We thus demonstrate efficient confinement of the magnon modes in perpendicularly-magnetized Bi-doped YIG (BiYIG) as well as the formation of  hybrid cavity-boundary modes. We show that when the anti-damping SOT is strong enough to overcome magnetic dissipation and drive one of the cavity boundaries into a deep nonlinear (incoherent) regime, the magnon confinement is disrupted. 
This happens because the boundary can no longer sustain stable dynamics and loses coherence, resulting in the extinction of the quantized cavity modes and their coupling with the boundary mode.
Relying on micromagnetic simulations, we identify cavity-boundary exchange coupling as the mode hybridization mechanism. Measurements conducted across a wide range of microwave excitation powers demonstrate further control of the coupled modes and mode anti-crossing.

The 3-nm-thick BiYIG film was grown by high-temperature RF magnetron sputtering on a (111)-oriented Y$_3$Sc$_2$Ga$_3$O$_{12}$ (YSGG) substrate [see Supporting Information (SI) for more detail]~\cite{legrand2024lattice}. Figure~\ref{fig1}(a) illustrates the device, where a stripline antenna, $\approx$~500 nm wide, is fabricated on the upper surface of BiYIG to excite propagating spin waves, with the wavevector direction along the $x$ axis. Additionally, two 7-nm-thick Pt nanostripes, each $\approx$~450~nm wide, are deposited near the antenna, separated by a center-to-center distance of $s=$ 2~$\upmu$m, leading to a gap between the Pt nanostripes of $a=$~1.55~$\upmu$m. These stripes were patterned using e-beam lithography, DC sputtering, and lift-off. Pt is known to induce additional interfacial anisotropy in garnets~\cite{lee2023large,lee2020interfacial}. Before measuring the fabricated devices, we used ferromagnetic resonance (FMR) to characterize the magnetic properties of an extended BiYIG film, both with and without Pt capping. Broadband FMR is measured on a microwave coplanar waveguide, with the external magnetic field in the film plane (resonance traces in SI). As shown in Fig.~\ref{fig1}(b), at any excitation frequency, the resonance field of BiYIG capped with Pt is reduced compared to the uncapped BiYIG, confirming the additional easy-plane anisotropy induced by Pt~\cite{lee2023large,lee2020interfacial}. To obtain the resulting effective magnetic anisotropies, we fit the resonance fields to the Kittel formula, $\omega=\gamma \mu_0 \sqrt{H_{\mathrm{ext}}\left(H_{\mathrm{ext}}+M_{\mathrm{eff}}\right)}$, where $M_{\mathrm{eff}}=M_{\rm s}+H_{\rm ani}$ is the effective magnetization, $H_{\rm ani}$ is negative for out-of-plane anisotropy and positive for easy-plane anisotropy, and $\gamma$ is the gyromagnetic ratio. The lines in Fig.~\ref{fig1}(b) show the fits, giving $\gamma/2\pi\approx$~28.5 $\pm$ 0.2~GHz/T and 28.4 $\pm$ 0.1~GHz/T, and $\mu_0M_{\rm eff}\approx$ 57 $\pm$ 23~mT and 9 $\pm$ 7~mT for BiYIG with and without Pt capping, respectively. The enhanced magnetic anisotropy of the BiYIG capped with Pt results in a locally modified magnon band structure, and thus forms a magnon cavity between the two Pt stripes, as shown schematically in Fig.~\ref{fig1}(a). We refer to the areas beneath the Pt stripes as the cavity boundaries. In addition to the increased anisotropy induced by the Pt capping, the Gilbert damping also rises from 0.0026 $\pm$ 0.0004 in the uncapped BiYIG film to 0.011 $\pm$ 0.001 in the Pt-capped sample (see SI). This increase is due to additional dissipation induced by spin pumping in Pt, which broadens the resonance linewidth of the modes in capped BiYIG~\cite{tserkovnyak2002}.

\begin{figure}
\hspace{-0.5cm}\includegraphics[width=160mm]{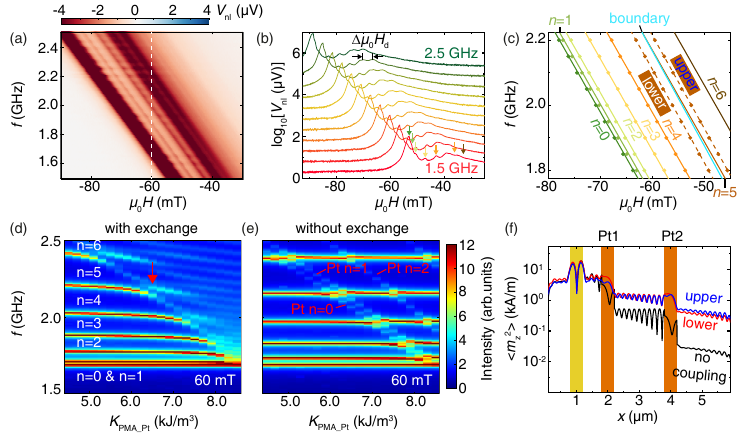}
\caption{(a) Spin pumping spectra measured in the Pt2 nanostripe at +15~dBm microwave power. (b) Line cuts extracted from (a) at different frequencies, representing the spin pumping voltage on a log scale. (c) Resonance fields of each cavity mode, extracted at different frequencies from (a) and (b). The solid lines are fits based on dipole-exchange spin-wave dispersion, the light blue curve corresponds to the boundary mode, and the brown curve is the predicted $n=5$ cavity mode. The brown dashed curves are theoretical fits of the hybridized modes. (d)-(e) Simulated resonance spectra of cavity and boundary modes at 60~mT as a function of PMA of the BiYIG beneath the Pt stripes, respectively with and without hybridization due to exchange coupling between capped and uncapped at the cavity boundary. The PMA-dependent resonances correspond to the quantized modes in the boundary region (BiYIG beneath the Pt stripe), denoted as Pt \(n=0, 1, 2\). (f) Simulated spatial profiles of the upper (blue curve) and lower (red curve) hybridized modes, and of the $n=5$ cavity mode without hybridization (black curve).}
\label{fig2}
\end{figure}

Next, we performed nonlocal spin pumping measurements on the Pt2 stripe.
A microwave current $I_{\rm rf}$ fed into the excitation antenna [Fig.~\ref{fig1}(a)] generates an oscillating magnetic field, which excites magnons that subsequently propagate a distance of 4~$\upmu$m to the Pt2 detector. At the detection site, a magnon-mediated spin current is emitted by spin-pumping and is locally converted into a charge current via the inverse spin Hall effect. The Pt2 stripe is connected to a lock-in amplifier to measure the resulting spin-pumping voltage ($V_{\mathrm{nl}}$).
The external magnetic field is applied along the $-x$ axis ($\varphi$ = 90$^\circ$) to maximize pumping efficiency. The microwave power is fixed at +15~dBm. Pulse modulation of the microwave source combined with a lock-in amplifier is used to improve the signal-to-noise ratio. Figure~\ref{fig1}(c) shows the spin-pumping voltage spectra, where two modes are observed, corresponding to the resonances of the uncapped BiYIG and cavity boundaries. The dashed lines represent fits based on Kittel's formula and the gyromagnetic ratio estimated from FMR. We find an effective anisotropy $\mu_0M_{\rm eff}=39\pm4$~mT and $4\pm2$~mT for the boundary and uncapped modes, respectively. These values are consistent with those obtained from FMR in Fig.~\ref{fig1}(b). The signal observed near zero field results from parametric pumping~\cite{wang2024broad,sandweg2011spin}. In this process, an incoming microwave photon excites pairs of counter-propagating magnons with reduced frequency and finite wave vectors, in accordance with both energy and momentum conservation.

Careful investigation of the spin pumping spectra reveals additional modes between the uncapped and boundary modes. They appear sharp in a finer scan within the frequency range 1.5-2.5 GHz, shown in Fig.~\ref{fig2}(a). Line cuts extracted from the spectra at different frequencies are plotted on a logarithmic scale in Fig.~\ref{fig2}(b). 
These regularly spaced resonances are associated to the standing wave patterns of the cavity formed between the two Pt stripes. The resonance fields of each mode (indicated by arrows of different colors) are extracted from the data at each frequency and shown in Fig.~\ref{fig2}(c).
We estimate the resonance fields using the dipole-exchange spin wave dispersion (see fitting details in SI)~\cite{dewames1970dipole,kalinikos1986theory,yu2014magnetic,qin2018propagating,wang2020chiral}.
The solid lines in Fig.~\ref{fig2}(c) are fits considering magnon dispersion and confinement, showing excellent agreement with the frequency of the $n=0-4$ modes when using for a cavity of length $a=$~1.55~$\upmu$m. 
In the absence of magnon-magnon coupling, the frequency of the $n$=5 cavity mode (brown line) is close to that of the boundary mode (blue line). Magnon-magnon coupling, however, causes mode hybridization with field splitting $\Delta\mu_0 H_d$, as observed in Figs.~\ref{fig2}(b,c). The frequency of the hybridized magnon branches can be modeled by considering the dipole-exchange spin-wave dispersion theory, Kittel's relation for the boundary mode, and the avoided crossing in a system of coupled resonators~\cite{Yuan2022}:
\begin{equation}
f_{\pm}=\frac{f_{\rm n}+f_{\rm b}}{2} \pm \sqrt{\left(\frac{f_{\rm n}-f_{\rm b}}{2}\right)^2+g^2},
\label{cross}
\end{equation}
where $f_{\rm n}$ and $f_{\rm b}$ are the quantized cavity and boundary mode frequencies, respectively, and $g$ denotes their mutual coupling strength. By adopting a coupling strength $g\approx$~0.051~GHz, the experimentally observed hybridized modes are reproduced, as shown by the two brown dashed curves in Fig.~\ref{fig2}(c).

To pinpoint the coupling mechanism between the boundary and cavity modes, we use MuMax$^3$ for micromagnetic simulations~\cite{vansteenkiste2014mumax3}. Our observations are consistent with a coupling mediated by the cavity-boundary exchange coupling, as shown by the comparison of Figs.~\ref{fig2}(d) and~\ref{fig2}(e), and magnon dynamics dominated by exchange interactions, as expected due to the negligible dipolar interaction in ultrathin BiYIG (see SI for details). 
The simulations also allow us to visualize the spatial profiles of the cavity modes excited by the microwave antenna, with and without hybridization with the boundary mode. The normalized time-averaged spatial intensities for both scenarios are shown in Fig.~\ref{fig2}(f). Without hybridization (black curve), the intensity distribution within the cavity is nearly uniform, with minimal leakage into the surrounding regions. However, once the pure cavity mode couples with the boundary mode (blue and red curves), the magnon transmission through the Pt1 boundary increases significantly. This coupling further results in non-uniform magnon intensity within the cavity due to leakage through Pt2.

\begin{figure}
\hspace{-0.5cm}\includegraphics[width=120mm]{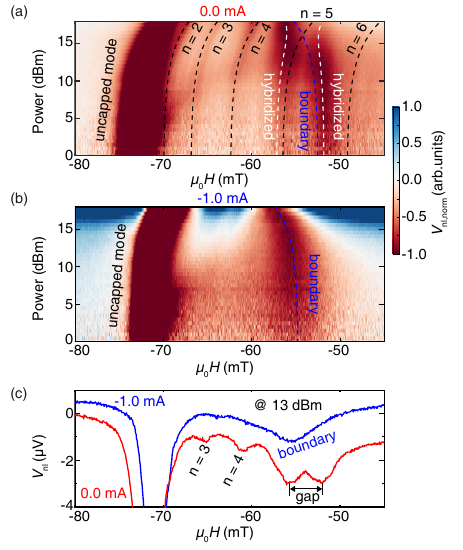}
\caption{Microwave power dependence of the spin pumping spectra without (a) and with an applied DC current of -1.0~mA (b). The excitation frequency is fixed at 2.0 GHz. At each power, the pumping voltages are normalized to the maximum amplitude of the boundary mode. The dashed lines in (a) are fits based on the magnon dispersion and Eq.~\ref{cross}. (c) Line cuts extracted from the raw data in (a) and (b) at a microwave power of +13~dBm, highlighting the anticrossing gap.} 
\label{fig3}
\end{figure}

To corroborate the coupling of the two modes, we control their detuning by using the frequency shift at high microwave powers~\cite{lee2023nonlinear}.
Figure~\ref{fig3}(a) shows spin-pumping spectra with the excitation frequency fixed at 2.0~GHz and the microwave power varied from 0 to +18~dBm. 
The quantized cavity modes are clearly observable at each power value. 
The microwave-induced heating and large precession cone angles at high powers cause the different modes to disperse as a function of magnetic field, as expected. The shifts of the resonance fields of the uncapped and boundary modes are opposite. This fact originates from the temperature altering not only  $M_\mathrm{s}$, but also the strain-induced crystalline anisotropy for the uncapped mode as well as the interfacial anisotropy for the boundary mode~\cite{maierflaig2017perpendicular}. To quantitatively assess the changes in the effective magnetization $M_{\rm eff}$ of these two modes, we used the dispersion and Eq.~\ref{cross} to reproduce the resonance field shifts of the two modes by considering a coefficient ($\zeta_P$) of evolution for $M_{\mathrm{eff}}$ due to the microwave power $P$~\cite{dzyapko2016lens,cherepanov1993saga}, $\mu_0M_{\mathrm{eff}} (P)=\mu_0M_{\mathrm{\rm eff}}^{\rm RT}+\zeta_p P,$ where the effective magnetization at room temperature for uncapped and boundary modes at zero power can be estimated using the values previously obtained at +15~dBm, respectively. The black and blue dashed curves in Fig.~\ref{fig3}(a) represent the fits for the cavity mode $n=5$ and the boundary mode before their hybridization, respectively. They intersect around +15~dBm, as discussed earlier in Fig.~\ref{fig2}, resulting in a distinct mode anti-crossing. The white dashed curves represent their corresponding hybridized modes (see Eq.~\ref{cross}) with $g \approx$ 0.06~$\pm$~0.01 GHz, in line with the value estimated from Fig.~\ref{fig2}. From these fits, we extract $\zeta_P\approx$~-0.12 mT/mW for the boundary mode and +0.09 mT/mW for the cavity modes. A positive $\zeta_P$ indicates that, in uncapped BiYIG, the heating-induced change in perpendicular magnetic anisotropy (PMA) is stronger than the reduction in saturation magnetization $M_\mathrm{s}$. In contrast, the negative $\zeta_P$ for the boundary mode likely results from a combination of the reduced $M_\mathrm{s}$, similar PMA changes in BiYIG, and additional modifications of the interfacial anisotropy due to the Pt capping. Note that the cavity modes exhibit low dissipation and are primarily confined between Pt1 and Pt2, resulting in weak spatial overlap with the Pt2 region where the spin-pumping signal is detected. Consequently, their spectral intensity is relatively low compared to the boundary modes, which dominate the signal due to stronger damping and greater spatial overlap with the detector, similar to the case of invisible magnon modes in magnon-polariton studies~\cite{zhang2014strongly}. The linewidths of both coupled modes exceed the coupling strength $g$. Nonetheless, the observed anticrossing features are clear signatures of hybridization and align well with both theoretical predictions and micromagnetic simulations. Additional evidence for the anti-crossing gap and mode hybridization is provided in the line cuts shown in Fig.~\ref{fig3}(c) and logarithmic plots of the normalized spin-pumping voltages (SI).

\begin{figure}
\hspace{-0.5cm}\includegraphics[width=120mm]{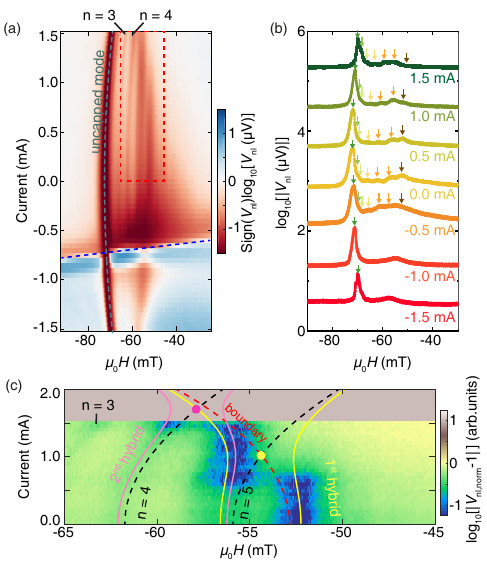}
\caption{(a) Spin pumping spectra measured in Pt2 at 2~GHz while applying different DC currents in Pt1 to modify the cavity and boundary modes. The green line are the resonance fields of uncapped mode as a function of DC current. The dashed blue line identifies the damping compensation condition, driven by spin injection in Pt1. (b) Line cuts extracted from (a) for DC currents from +1.5 to -1.5~mA in steps of 0.5~mA. The arrows indicate the different cavity modes. (c) Detail of the spectra in the dashed rectangle in (a), highlighting the mode anti-crossings at positive current. The dashed lines are fits of current-modified uncoupled cavity (black) and boundary (red) modes. The yellow lines show the hybridization between the $n=5$ and boundary modes (1$^{\rm st}$ hybridization), with the corresponding crossing point marked by a yellow dot. The purple lines indicate the hybridization between the $n=4$ and the 1$^{\rm st}$ hybridized mode (2$^{\rm nd}$ hybridization), with the crossing point highlighted by a purple dot.}
\label{fig4}
\end{figure}

In our system where the cavity boundaries are formed by the Pt/BiYIG heterostructures, not only Pt can act as an inverse spin Hall detector, but it can also serve as a modulator of magnon dissipation through the interfacial injection of spin-polarized currents. To explore the impact of SOTs on the cavity and boundary modes and their couplings, a DC current is applied to the central Pt1 stripe, injecting angular momentum into BiYIG via the spin Hall effect (SHE). 
A distinct difference is observed in the power-dependent measurements when a -1.0~mA DC current is applied to the Pt1 stripe [Fig.~\ref{fig3}(b)], compared to the case where SOTs are absent [Fig.~\ref{fig3}(a)]. To further explore the effect of spin current injection via Pt1, we fix the microwave power at +10.5~dBm to keep the cavity and boundary modes separated in the absence of current, and vary the magnitude and sign of the DC current, as shown in Fig.~\ref{fig4}.

A first effect of the current is thermal, similar to the microwave power dispersion presented above. As shown in Fig.~\ref{fig4}(a), the resonance fields of all modes exhibit symmetrical shifts with respect to zero current, consistent with a thermal effect. 
Figure~\ref{fig4}(b) shows linecuts of the spectra shown in (a) for different currents injected in Pt1. The dashed lines in Fig.~\ref{fig4}(a) and~\ref{fig4}(c) are the fits to the modes based on the following equation, $\mu_0M_{\mathrm{eff}} ( I)=\mu_0M_{\mathrm{\rm eff}}^{\rm RT}+\zeta_I I^2$, where $\zeta_I$ is the coefficient of evolution of $M_{\mathrm{eff}}$ due to the current $I$. From the fit, we find $\zeta_I\approx$~+2.36 and -4.00 mT/(mA)$^2$ for the uncapped and boundary modes, respectively.
By using the estimated $\zeta_I$ combined with the dispersion and Eq.~\ref{cross}, we reproduce all the observed hybridized modes in Fig.~\ref{fig4}(c). As the DC current increases from 0 to 1.5~mA, the $n$=5 cavity and boundary modes intersect around 1~mA (yellow dot), creating an anti-crossing gap also at +10.5 dBm. 
The yellow curves in Fig.~\ref{fig4}(c) represent the modes formed by hybridization of the $n$ = 5 cavity mode and the boundary mode calculated by assuming a coupling strength of $g_1\approx$~0.07~$\pm$~0.01~GHz. These modes, referred to as the 1$^{\rm st}$ hybridization, show good agreement with the experimental results for currents below 1~mA. However, as the current increases to 1.5~mA, the left branch of the 1$^{\rm st}$ hybridized mode deviates from the experimental spectra. This discrepancy arises because the left branch undergoes a second crossing with the $n$ = 4 cavity mode around 1.7~mA (purple dot), beyond the experimental limit imposed by Pt1 breakdown at currents above 1.5~mA. This 2$^{\rm nd}$ hybridization, between the left branch of the 1$^{\rm st}$ hybridized mode and the $n$ = 4 cavity mode, is required for precisely reproducing the observed mode dispersion. The resulting modes, shown as purple lines in Fig.~\ref{fig4}(c), match well with the experimental data. From the fit, we estimate a coupling strength of $g_2\approx$~0.05~$\pm$~0.01~GHz for this 2$^{\rm nd}$ hybridization. These results indicate that mode anti-crossings are not exclusive to the $n = 5$ mode and can occur for any cavity mode that intersects with the boundary mode.

Finally, we discuss the control of the cavity and boundary modes via spin current injection. A positive current ($I$ along +$y$) applied to Pt1 for external magnetic field towards -$x$ increases the damping, reducing the precession amplitude of both cavity and boundary modes. Their hybridization at large current remains noticeable (top of Figs.~\ref{fig4}(a) and~\ref{fig4}(c)), although with lower signals. In contrast, for $I$ along -$y$, the anti-damping SOT offsets and even overcomes the damping in Pt-capped BiYIG. Consequently, for increasing currents, the amplitude of the boundary mode and the corresponding pumping voltages are enhanced, until a threshold is reached at which the damping is fully compensated, indicated by the blue dashed line in Fig.~\ref{fig4}(a)~\cite{merbouche2024true}. The current required for damping compensation increases with magnetic field, explaining the slope of the threshold in Fig.~\ref{fig4}(a). Compensation of damping by SOTs leads to considerably larger populations of both coherent and incoherent magnons. Incoherent magnons diffuse in BiYIG, and are collected by the Pt2 detector, which generates a purely DC signal, magnified near the damping compensation~\cite{wimmer2019spin}. The observed sign change of the spin-pumping background signal across the threshold is induced by the modulation of the DC signal (see detailed discussion in SI). Since this signal is related to magnons created by the DC current in Pt1, it is expected to be independent of the coherent spin waves excited by the microwave antenna. We verify this aspect by a control experiment in which the excitation frequency of the microwave antenna is varied, which reveals no effect on the damping compensation signal (see SI for details). 

When the current is further increased beyond damping compensation, nonlinear processes dominate and magnons may even enter a chaotic regime, leading to a loss of the spatial coherence and to a significant suppression of the spin-pumping voltage~\cite{merbouche2024true}. In addition, the BiYIG below Pt1, acting as one of the cavity boundaries, does not present a homogeneous magnetic state, thus disrupting the cavity formation. The previously observed cavity modes and their coupling with the boundary mode are no longer present, as evidenced by the suppression of their multiple peaks in Fig.~\ref{fig4}(b). Correspondingly, only the uniform modes of the BiYIG, both with and without Pt capping, remain observable.
Therefore, by varying the injector current, we can effectively turn the cavity modes on and off and control their magnon-magnon coupling with the boundary mode, a feature which works at any microwave excitation power. As shown in Fig.~\ref{fig3}(b), for a sufficiently large injector current beyond damping compensation, such as -1.0~mA, the cavity modes, and consequently their coupling to the boundary mode, are effectively suppressed.

In summary, our observations highlight the potential of magnonic nano-resonators as a promising platform to build magnonic devices. First, we demonstrated the use of the interfacial anisotropy induced by two Pt nanostripes to create an on-chip magnonic cavity, offering an alternative to previous works employing interlayer dipolar couplings~\cite{santos2023magnon,qin2021nanoscale,vogel2015optically,hong2021tunable}. Second, we used the inverse spin Hall voltage generated by spin pumping in Pt as a powerful method to probe the quantized cavity and boundary modes within a nanostructured magnonic system. The different modes of the magnonic device were successfully modeled based on dipole-exchange spin-wave dispersion. We estimated the coupling strength between the cavity and boundary modes and assigned it to the cavity-boundary exchange interaction, corroborated by micromagnetic simulations. Finally, we demonstrated that spin injection in the cavity boundaries allows for in-situ control over the cavity formation and its interaction with boundary modes. Our results open new possibilities for further exploration of magnon lasers~\cite{malz2019topological} and cascaded logic devices~\cite{manipatruni2019scalable}, for which the possibility to dynamically reconfigure magnonic properties and coupling between modes is essential. The compact footprint and planar geometry of our design further allow for the incorporating multiple magnonic cavities within a small area. This would enable engineered coupling between adjacent cavities, opening a path toward complex magnonic networks and on-chip coherent magnonic circuits.

%

\begin{acknowledgement}

This research was supported by the Swiss National Science Foundation (Grant No. 200020-200465). H.W. acknowledges the support of the China Scholarship Council (CSC, Grant No. 202206020091). W.L. acknowledges the support of the ETH Zurich Postdoctoral Fellowship Program (21-1 FEL-48).

\end{acknowledgement}

\begin{suppinfo}

The data underlying this study are openly available in ETH Research Collection~\cite{Data_ETH}. 

The Supporting Information is available free of charge at [link],

BiYIG sample growth and characterization, topography of the device measured by AFM, rationale for using the 3-nm-thick BiYIG thin film, SQUID measurement of the saturation magnetization of BiYIG, spin-wave dipole-exchange dispersion and the fitting parameters to the cavity modes, micromagnetic simulations of the anticrossings, frequency dependence of the observed signal at the threshold current for damping compensation, explanation of the sign change across the damping compensation conditions, spin pumping voltage spectra measured under both positive and negative fields, and power-dependent spin pumping spectra with improved visibility.

\end{suppinfo}


\end{document}